\documentclass[twocolumn,amsmath,amssymb,aip,pop]{revtex4}

\usepackage{graphicx}% Include figure files
\usepackage{bm}% bold math
\usepackage{amsfonts}

\usepackage[utf8]{inputenc}

\begin{document}
\title{Nonlinear theory of the modulational instability at the ion-ion hybrid frequency
and collapse of ion-ion hybrid waves in two-ion plasmas}
\author{Volodymyr M. Lashkin}
\email{vlashkin62@gmail.com} \affiliation{$^1$Institute for
Nuclear Research, Pr. Nauki 47, Kyiv 03028, Ukraine}
\affiliation{$^2$Space Research Institute, Pr. Glushkova 40 k.4/1,
Kyiv 03187,  Ukraine}

%\date{\today}

\begin{abstract}
We study the dynamics of two-dimensional nonlinear ion-ion hybrid
waves propagating perpendicular to an external magnetic field in
plasmas with two ion species. We derive nonlinear equations for
the envelope of electrostatic potential at the ion-ion hybrid
frequency to describe the interaction of ion-ion hybrid waves with
low frequency acoustic-type disturbances. The resulting nonlinear
equations also take into account the contribution of second
harmonics of the ion-ion hybrid frequency. A nonlinear dispersion
relation is obtained and, for a number of particular cases, the
modulational instability growth rates are found. By neglecting the
contribution of second harmonics, the phenomenon of collapse of
ion-ion hybrid waves is predicted. It is shown that taking into
account the interaction with the second harmonics results in the
existence of a stable two-dimensional soliton.
\end{abstract}

\maketitle

\section{Introduction}

The presence of several species of ions is often found in both
space and laboratory plasmas. In particular, space plasmas in most
cases consist of several species of ions and the relative
concentration of different species can vary in a fairly wide
range. For example, the ionospheric and plasmaspheric plasma are
composed of several species of ions \cite{Horvitz1990}, and in the
upper ionosphere $\mathrm{O}^{+}$ ions with a small addition of
$\mathrm{He^{+}}$ are predominant. Phenomena in multi-ion space
plasmas have been intensively studied for many years
\cite{Ashour1988,Lee2008,Wang2012,Bordikar2013}. In laboratory
conditions, a plasma with two ion species is of great interest
primarily in relation to the ion cyclotron resonance frequency
(ICRF) heating method in plasma magnetic confinement devices,
where one of the most successful schemes involves minority species
heating at the ion-ion hybrid resonance or at the minority
cyclotron frequency \cite{Adam1987,Porkolab1998,Oliver2014} in
$\mathrm{H}-\mathrm{D}$ plasma. Recently, efficient plasma heating
with the three-ion ICRH scenario with a small amount of
$^{3}\mathrm{He}$ ions in $\mathrm{H}-\mathrm{D}$ mixture was
suggested in Ref. \cite{Kazakov2015,Kazakov2021}. The presence of
two ion species is inherent to dusty plasmas
\cite{Rao1990,Shukla2001,Fortov2005}, where in addition to the
main ion component, a second, heavy micronsize ions are present.
Two-ion plasmas, although unmagnetized, also naturally arises in
the inertial thermonuclear fusion experiments \cite{Simakov2018}.

In a magnetized plasma consisting of electrons and two ion species
with different charge-to-mass ratios, in addition to the lower-
and upper-hybrid resonances, there is the so-called ion-ion hybrid
resonance at the frequency $\omega_{ii}$ defined by
\begin{equation}
\label{hybrid-frequency}
\omega_{ii}^{2}=\frac{\omega_{p1}^{2}\Omega_{2}^{2}
+\omega_{p2}^{2}\Omega_{1}^{2}}{\omega_{p1}^{2}+\omega_{p2}^{2}}
=\frac{(m_{1}n_{01}+m_{2}n_{02})Z_{1}Z_{2}\Omega_{1}\Omega_{2}}{m_{2}n_{01}Z_{1}^{2}
+m_{1}n_{02}Z_{2}^{2}},
\end{equation}
and first introduced by Buchsbaum \cite{Buchsbaum2}. Here,
$\omega_{p\alpha}$ and $\Omega_{\alpha}$ are the plasma frequency
and gyrofrequency of the ions of species $\alpha=1,2$,
respectively with $\omega_{p\alpha}^{2}=4\pi Z_{\alpha}^{2}
e^{2}n_{0\alpha}/m_{\alpha}$ and
$\Omega_{\alpha}=Z_{\alpha}eB_{0}/m_{\alpha}c$, where $e$ is the
elementary charge, $n_{0\alpha}$, $m_{\alpha}$ and $Z_{\alpha}$
are the equilibrium density, the mass and charge number of the
ions of species $=1,2$, respectively. Overall charge neutrality
$n_{01}+n_{02}=n_{0e}=n_{0}$ is assumed, where $n_{0}$ is the
equilibrium plasma density and $n_{0e}$ is the equilibrium
electron density. From Eq. (\ref{hybrid-frequency}) one can see
that the ion-ion hybrid frequency $\omega_{ii}$ lies between the
gyrofrequencies $\Omega_{1}$ and $\Omega_{2}$ of the ions of
different species. Note also that $\omega_{ii}$ is determined only
by the magnetic field $B_{0}$ and the relative population of each
ion species. The presence of an additional type of ions in a
magnetized plasma significantly modifies the dispersion relation
and leads to the appearance of new branches of plasma oscillations
that are absent in the single-ion case. The properties of such a
plasma differ in many respects from the properties of single-ion
plasma. Linear theory of wave propagation in plasmas with two
species of ions, including an inhomogeneous plasma, has been
considered in quite a few works (see, e.g.,
Refs.~\cite{Brice1964,Swanson1976,Perkins1977,Gendrin1984,Brambilla1985,Mann1997}).
Parametric instabilities in a plasma with two ion species were
investigated in
Refs.~\cite{Kitzenko1973,Kaw1973,Ott1973,Porkolab1980}, where the
standard kinetic method for studying parametric instabilities
\cite{Silin1965,Nishikawa1968} was used, as well as in
Refs.~\cite{Lee1975,Chan1975} within the framework of fluid model.

The linear theory is valid only for sufficiently small wave
amplitudes, when nonlinear effects can be neglected. Nonlinear
coherent structures in plasma, in particular solitons, have been
the subject of intensive theoretical study for several decades and
have been experimentally observed both in laboratory and space
plasmas \cite{Petviashvili_book1992,Horton1996,Scorich2010}. In a
broad sense, a soliton is a localized structure (not necessarily
one-dimensional) resulting from the balance of dispersion and
nonlinearity effects. Multidimensional solitons often turn out to
be unstable, and the most well-known phenomena in this case are
wave collapse and wave breaking
\cite{Kuznetsov1986,KuznetsovZakharov2000,Zakharov_UFN2012}.
Despite the obvious importance of studying nonlinear phenomena
occurring in a plasma with two species of ions in the vicinity of
the ion-ion hybrid  frequency, the corresponding nonlinear theory,
especially in the multidimensional case, has not been sufficiently
developed, in contrast to the lower-hybrid (LH) and upper-hybrid
(UH) resonances, for which nonlinear phenomena have been studied
in more detail. In particular, in the one-dimensional case,
various types of solitons, including envelope solitons, were
discovered at the LH \cite{Yu1978,Jovanovic2007, Saleem2022} and
UH frequencies \cite{Stenflo1975,Goldman1976,Shukla1977}. In the
multidimensional case, the phenomenon of collapse of the LH
\cite{Sturman1976,Shapiro1984,Shapiro1993,Shapiro1995,Shukla2005}
and UH waves \cite{Rasmussen1983,Shukla1984} was predicted. Taking
nonlocal nonlinearity into account, stable two-dimensional UH
solitons and vortex solitons were found in
Ref.~\cite{Lashkin2007}.

Note that one-dimensional solitons in a plasma with two species of
ions were considered in a number of works
\cite{Lashkin1990,Mikhailovskii1985,Irie2001,Wang2002,Irie2003,
Chaudhuri2005,Verheest2012,Rufai2019,Kaladze2019}, but they were,
with the exception of Refs.~\cite{Lashkin1990,Wang2002}, not the
envelope solitons at the ion-ion hybrid frequency (that is, they
were not ion-ion hybrid solitons), but to solitons in the
frequency range of much lower or much higher the ion-ion hybrid
frequency $\omega_{ii}$.

One-dimensional (1D) nonlinear waves near the ion-ion hybrid
frequency $\omega_{ii}$ were considered in
Refs.~\cite{Lashkin1990,Wang2002}. In both of those works,
equations  for the wave envelope at the ion-ion hybrid frequency
$\omega_{ii}$  were derived. In Ref.~\cite{Lashkin1990}, an
equation with nonlocal nonlinearity was obtained, however, the
nonlinearity was incorrectly taken into account due to the neglect
of the usual striction nonlinearity associated with the
ponderomotive force, in comparison with the nonlocal nonlinearity
due to the interaction with the second harmonics. In
Ref.~\cite{Wang2002}, the 1D nonlinear Schr\"{o}dinger equation
(of both focusing and defocusing types) was obtained. In that
work, however, the action of the ponderomotive force of the HF
field of ion-ion hybrid waves on ions was completely neglected,
although the ion contribution is comparable (and sometimes
exceeds) the electron contribution. Besides, the dispersion of
low-frequency waves was incorrectly taken into account, so that
the results obtained in Ref.~\cite{Wang2002} have a very limited
area of applicability. We also note that in the one-dimensional
theory there is no possibility of taking into account the vector
(gyrotropic) nonlinearity, similar to the nonlinearity that occurs
near the LH resonance \cite{Sturman1976,Shapiro1993}.

The aim of this paper is to obtain two-dimensional (2D) nonlinear
equations to describe the interaction of ion-ion hybrid waves with
low-frequency (LF) acoustic-type disturbances. In the equation for
the envelope we also take into account the interaction of second
harmonics at the ion-ion hybrid  frequency. In the case where
second harmonics can be neglected, we predict a collapse of
ion-ion hybrid waves, similar to the collapse for the LH and UH
waves. Taking into account the additional nonlinearity associated
with the second harmonic, however, leads to a stable 2D soliton.

The paper is organized as follows. In Sec. II, we derive a set of
nonlinear equations for the wave envelope and LF ion density
perturbations. A nonlinear dispersion relation was obtained in
Sec. III. In Sec. IV the phenomenon of collapse of ion-ion hybrid
waves is predicted. A stable 2D soliton, taking into account the
second harmonic, was found in Sec. V. Finally, Sec. VI concludes
the paper.

\section{\label{sec2} Model equations}

For a cold plasma containing two ion species and immersed in a
homogeneous external magnetic field
$\mathbf{B}_{0}=B_{0}\mathbf{\hat{z}}$, where $\mathbf{\hat{z}}$
is the unit vector along the $z$-direction, the linear dispersion
relation for electrostatic waves propagating normal to
$\mathbf{B}_{0}$ is given by \cite{Stix1992}
\begin{equation}
\label{disp-0} 1-\frac{\omega_{pe}^{2}}{\omega^{2}-\Omega_{e}^{2}}
-\frac{\omega_{p1}^{2}}{\omega^{2}-\Omega_{1}^{2}}
-\frac{\omega_{p2}^{2}}{\omega^{2}-\Omega_{2}^{2}}=0,
\end{equation}
where $\omega_{pe}$ and $\Omega_{e}$ are the electron plasma
frequency and electron-cyclotron frequency, respectively. In the
high frequency range $\omega\gg\omega_{p1,2},\Omega_{1,2}$, where
only electrons take part in the plasma motion, for the
upper-hybrid frequency $\omega_{UH}$ we get,
\begin{equation}
\omega_{UH}^{2}=\omega_{pe}^{2}+\Omega_{e}^{2}.
\end{equation}
In the intermediate frequency range
$\Omega_{1},\Omega_{2}\ll\omega\ll\omega_{pe},\Omega_{e}$,
solution of Eq. (\ref{disp-0}) yields the frequency of the
lower-hybrid resonance $\omega_{LH}$,
\begin{equation}
\omega_{LH}^{2}=\frac{\omega_{p1}^{2}+\omega_{p2}^{2}}{1+\omega_{pe}^{2}/\Omega_{e}^{2}}.
\end{equation}
Here, only ions play an active role in motion (the role of
electrons is reduced to screening). Assuming $\omega\ll
\Omega_{e}$, equation (\ref{disp-0}) can be written as
\begin{gather}
\omega^{4}-\omega^{2}\left[\Omega_{1}^{2}+\Omega_{2}^{2}
+\frac{\omega_{p1}^{2}+\omega_{p2}^{2}}{1+\omega_{pe}^{2}/\Omega_{e}^{2}}\right]
\nonumber \\
+\Omega_{1}^{2}\Omega_{2}^{2}
+\frac{\omega_{p1}^{2}\Omega_{2}^{2}+\omega_{p2}^{2}\Omega_{1}^{2}}{1+\omega_{pe}^{2}/\Omega_{e}^{2}}=0.
\label{aux1}
\end{gather}
In the lowest frequency range
$\Omega_{1},\Omega_{2}\ll\omega_{p1},\omega_{p2}$, we have
\begin{equation}
\Omega_{1}^{2}+\Omega_{2}^{2}\ll
\frac{\omega_{p1}^{2}+\omega_{p2}^{2}}{1+\omega_{pe}^{2}/\Omega_{e}^{2}}
\quad \mathrm{and}\quad \Omega_{1}^{2}\Omega_{2}^{2}\ll
\frac{\omega_{p1}^{2}\Omega_{2}^{2}+\omega_{p2}^{2}\Omega_{1}^{2}}{1+\omega_{pe}^{2}/\Omega_{e}^{2}},
\end{equation}
and Eq. (\ref{aux1}) can be simplified to
\begin{equation}
\label{disp-ion11}
\omega^{4}-\omega^{2}\frac{(\omega_{p1}^{2}+\omega_{p2}^{2})}{(1+\omega_{pe}^{2}/\Omega_{e}^{2})}
+\frac{\omega_{p1}^{2}\Omega_{2}^{2}+\omega_{p2}^{2}\Omega_{1}^{2}}{1+\omega_{pe}^{2}/\Omega_{e}^{2}}=0.
\end{equation}
On assuming that
\begin{equation}
\omega^{2}\ll
\frac{\omega_{p1}^{2}+\omega_{p2}^{2}}{1+\omega_{pe}^{2}/\Omega_{e}^{2}},
\end{equation}
equation (\ref{disp-ion11}) yields the ion-ion hybrid frequency
$\omega_{ii}$ determined by Eq. (\ref{hybrid-frequency}). In this
case,
\begin{equation}
\label{disp-ion} \frac{\omega_{p1}^{2}}{\omega^{2}-\Omega_{1}^{2}}
+\frac{\omega_{p2}^{2}}{\omega^{2}-\Omega_{2}^{2}}=0.
\end{equation}
Ions of both species equally take part in the plasma motion and
move in opposite phases, and it is this, compared with the cases
of UH and LH resonances, that complicates the description of the
behavior of electrostatic waves near the ion-ion hybrid resonance.

In this section, we derive nonlinear equations to describe the
dynamics of waves near the ion-ion hybrid frequency $\omega_{ii}$.
We consider the case of an arbitrary ratio of the ion densities
$n_{01}$ and $n_{02}$, as well as arbitrary ratios of the electron
and ion temperatures. The reason is that in plasma of the Earth's
ionosphere, for example, the ratios of these quantities can be,
depending on the altitude, either comparable to each other, or
significantly (sometimes by orders of magnitude) differ from each
other in one side or another \cite{Horvitz1990}.

\subsection{High-frequency disturbances}

The basic equations governing the plasma dynamics are the fluid
equations of motion and continuity of the ions of both species and
electrons,
\begin{equation}
\label{motion_eq} \frac{\partial \mathbf{v}_{\alpha}}{\partial
t}+\left(\mathbf{v}_{\alpha}\cdot\nabla\right)
\mathbf{v}_{\alpha}=-\frac{e_{\alpha}}{m_{\alpha}}\nabla\varphi-\frac{\nabla
p_{\alpha}}{m_{\alpha}n_{\alpha}}+\Omega_{\alpha}[\mathbf{v}_{\alpha}\times
\hat{\mathbf{z}}],
\end{equation}
\begin{equation}
\label{continuity_eq} \frac{\partial n_{\alpha}}{\partial
t}+\nabla\cdot (n_{\alpha}\mathbf{v}_{\alpha})=0,
\end{equation}
where $n_{\alpha}$, $\mathbf{v}_{\alpha}$, $p_{\alpha}$,
$e_{\alpha}$ and $m_{\alpha}$ are the density, velocity, pressure,
charge, and mass of the particle species $\alpha=e,i$ (electrons
and ions), respectively. For the gas kinetic pressure we take
$p_{\alpha}=\gamma_{\alpha}n_{\alpha}T_{\alpha}$, where
$T_{\alpha}$ is the temperature and $\gamma_{\alpha}$ is the ratio
of specific heats, and in the next we introduce the notation
$v_{T\alpha}=\sqrt{\gamma_{\alpha}T_{\alpha}/m_{\alpha}}$ for the
particle thermal velocity. Equations (\ref{motion_eq}) and
(\ref{continuity_eq}) are supplemented by the Poisson equation,
\begin{equation}
\label{Poisson} \Delta\varphi=4\pi (en_{e}-e_{1}n_{1}-e_{2}n_{2}).
\end{equation}
Following the well-known idea dating back to the original work by
Zakharov \cite{Zakharov1972} of separating the slow and fast time
scales and averaging over the fast time, we represent the
electrostatic potential, velocities, and  densities in the form
\begin{equation}
\label{phi}
\varphi=\varphi^{(s)}+\left(\frac{1}{2}\varphi^{(1)}\mathrm{e}^{-i\omega_{ii}t}
+\varphi^{(2)}\mathrm{e}^{-2i\omega_{ii}t}+\mathrm{c.\,c.}\right),
\end{equation}
\begin{equation}
\label{v}
\mathbf{v}_{\alpha}=\mathbf{v}_{\alpha}^{(s)}+\left(\frac{1}{2}\mathbf{v}_{\alpha}^{(1)}\mathrm{e}^{-i\omega_{ii}t}
+\mathbf{v}_{\alpha}^{(2)}\mathrm{e}^{-2i\omega_{ii}t}+\mathrm{c.\,c.}\right),
\end{equation}
\begin{equation}
\label{n} n_{\alpha}=n_{0\alpha}+\delta
n_{\alpha}+\left(\frac{1}{2}n_{\alpha}^{(1)}\mathrm{e}^{-i\omega_{ii}t}
+n_{\alpha}^{(2)}\mathrm{e}^{-2i\omega_{ii}t}+\mathrm{c.\,c.}\right),
\end{equation}
where c.c. stands for the complex conjugate,
$\mathbf{v}_{\alpha}^{(1),(2)}$ and $n_{\alpha}^{(1),(2)}$ are
assumed to vary on a timescale much more slowly than
$1/\omega_{ii}$. The contribution of the second harmonic of
frequency $\omega_{ii}$ is taken into account in Eqs. (\ref{phi}),
(\ref{v}) and (\ref{n}), and we assume that the amplitudes of the
first harmonic are much larger than the others so that conditions
$|\mathbf{v}_{\alpha}^{(1)}|\gg |\mathbf{v}_{\alpha}^{(2)}|,
|\mathbf{v}_{\alpha}^{(s)}|$ and $n_{\alpha}^{(1)}\gg
n_{\alpha}^{(2)}, \delta n_{\alpha}$ are assumed to be met. Note
that, as shown in Refs.
\cite{Kuznetsov1976,Scorich1980,Scorich2010}, taking into account
the second harmonic of the Langmuir frequency $\omega_{pe}$ in a
nonmagnetized plasma may halt Langmuir collapse in two or three
dimensions.

The ion-ion hybrid waves have wave numbers almost normal to the
external magnetic field ($k_{z}\ll k_{\perp}$), and in this paper
we restrict ourselves to the 2D case of perpendicular propagation
when the condition
\begin{equation}
\label{perp-cond} k_{z}^{2}/k_{\perp}^{2}\ll m_{e}/m_{1,2}
\end{equation}
is satisfied. Substituting Eq.~(\ref{v}) into
Eq.~(\ref{motion_eq}) , we have
\begin{gather}
 \left(-i\omega_{ii}+\frac{\partial}{\partial
t}\right)\mathbf{v}_{\alpha}^{(1)}+\mathbf{F}_{\alpha}^{(1)}=-\frac{e}{m_{\alpha}}\nabla_{\perp}\varphi^{(1)}
\nonumber \\
-\frac{v_{T\alpha}^{2}}{n_{0\alpha}}\nabla_{\perp}
n_{\alpha}^{(1)} +\Omega_{\alpha}[\mathbf{v}_{\alpha}^{(1)}\times
\hat{\mathbf{z}}], \label{v1}
\end{gather}
where
\begin{equation}
\mathbf{F}_{\alpha}^{(1)}=(\mathbf{v}_{\alpha}^{(1)\ast}\cdot\nabla_{\perp})\mathbf{v}_{\alpha}^{(2)}
+(\mathbf{v}_{\alpha}^{(2)}\cdot\nabla_{\perp})\mathbf{v}_{\alpha}^{(1)\ast},
\label{F}
\end{equation}
where $\nabla_{\perp}=(\partial/\partial x,\partial/\partial y)$.
Substituting Eq.~(\ref{n}) into Eq.~(\ref{continuity_eq}), we find
\begin{equation}
\label{n1} \left(-i\omega_{ii}+\frac{\partial}{\partial
t}\right)n_{\alpha}^{(1)}+n_{0\alpha}\nabla_{\perp}\cdot\mathbf{v}_{\alpha}^{(1)}
+\nabla_{\perp}\cdot\mathbf{J}_{\alpha}^{(1)}=0,
\end{equation}
where $\mathbf{J}_{\alpha}^{(1)}$ is the nonlinear current,
\begin{equation}
\label{j} \mathbf{J}_{\alpha}^{(1)}=\delta
n_{\alpha}\mathbf{v}_{\alpha}^{(1)}
+n_{\alpha}^{(1)}\mathbf{v}_{\alpha}^{(s)}
+n_{\alpha}^{(1)\ast}\mathbf{v}_{\alpha}^{(2)}
+n_{\alpha}^{(2)}\mathbf{v}_{\alpha}^{(1)\ast}.
\end{equation}
As noted above, electrons do not take part in the plasma motion at
the frequency of the ion-ion hybrid resonance, and ions of
different species move in opposite phases, so we can write
\begin{equation}
\label{Pois1} e_{1}n_{1}^{(1)}+e_{2}n_{2}^{(1)}=0.
\end{equation}
In zero order in
$i\omega_{ii}\partial_{t}/(\Omega_{\alpha}^{2}-\omega_{ii}^{2})\ll
1$, neglecting the thermal dispersion and nonlinearity, from Eq.
(\ref{v1}) we obtain the perpendicular velocity
\begin{equation}
\label{v0}
\mathbf{v}_{0\alpha}^{(1)}=\frac{ie_{\alpha}\omega_{ii}}{m_{\alpha}(\Omega_{\alpha}^{2}-\omega_{ii}^{2})}
\left(\nabla_{\perp}\varphi^{(1)}+i\frac{\Omega_{\alpha}}{\omega_{ii}}[\nabla_{\perp}\varphi^{(1)}
\times\hat{\mathbf{z}}]\right)
\end{equation}
and then from Eq. (\ref{n1})  the density perturbation
\begin{equation}
\label{n0}
n_{0\alpha}^{(1)}=\frac{n_{0\alpha}}{i\omega_{ii}}\nabla_{\perp}\cdot\mathbf{v}_{0\alpha}^{(1)}
=\frac{1}{4\pi
e_{\alpha}}\frac{\omega_{p\alpha}^{2}}{(\Omega_{\alpha}^{2}-\omega_{ii}^{2})}\Delta_{\perp}\varphi^{(1)},
\end{equation}
where $\Delta_{\perp}=\partial^{2}/\partial
x^{2}+\partial^{2}/\partial y^{2}$. In the following order, taking
into account the thermal dispersion and nonlinearity, one can
obtain
\begin{gather}
\mathbf{v}_{\alpha}^{(1)}=\frac{i\omega_{ii}}{\Omega_{\alpha}^{2}-\omega_{ii}^{2}}\left[1
+\frac{2i\omega_{ii}\partial_{t}}{(\Omega_{\alpha}^{2}-\omega_{ii}^{2})}\right]
\nonumber \\
\times\left\{\left[\left(1+\frac{i\partial_{t}}{\omega_{ii}}\right)\frac{e_{\alpha}}{m_{\alpha}}
\nabla_{\perp}\varphi^{(1)}
+\frac{v_{T\alpha}^{2}}{n_{0\alpha}}\nabla
n_{\alpha}^{(1)}+\mathbf{F}_{\alpha}^{(1)}\right]\right.
\nonumber \\
\left.
+\frac{i\Omega_{\alpha}}{\omega_{ii}}\left[\left(\frac{e_{\alpha}}{m_{\alpha}}\nabla_{\perp}\varphi^{(1)}
+\frac{v_{T\alpha}^{2}}{n_{0\alpha}}\nabla
n_{\alpha}^{(1)}+\mathbf{F}_{\alpha}^{(1)}\right)\times\hat{\mathbf{z}}\right]\right\},
\label{v1-full}
\end{gather}
where we have neglected non-stationary corrections $\sim
\partial_{t}\nabla
n_{\alpha}^{(1)}$ and $\sim
\partial_{t}\mathbf{F}^{(1)}$ in terms responsible for thermal
dispersion and nonlinearity, respectively. Using
Eq.~(\ref{v1-full}) we substitute
$\nabla_{\perp}\cdot\mathbf{v}_{\alpha}^{(1)}$ into Eq.~(\ref{n1})
and get
\begin{gather}
 \left(-i\omega_{ii}+\frac{\partial}{\partial
t}\right)n_{\alpha}^{(1)}+\frac{in_{0\alpha}\omega_{ii}}{\Omega_{\alpha}^{2}-\omega_{ii}^{2}}
\left[1
+\frac{2i\omega_{ii}\partial_{t}}{(\Omega_{\alpha}^{2}-\omega_{ii}^{2})}\right]
\nonumber \\
\times\left\{\left[\left(1+\frac{i\partial_{t}}{\omega_{ii}}\right)\frac{e_{\alpha}}{m_{\alpha}}
\Delta_{\perp}\varphi^{(1)}
+\frac{v_{T\alpha}^{2}}{n_{0\alpha}}\Delta
n_{\alpha}^{(1)}+\nabla_{\perp}\cdot\mathbf{F}_{\alpha}^{(1)}\right]\right.
\nonumber \\
\left.
+\frac{i\Omega_{\alpha}}{\omega_{ii}}\nabla_{\perp}\cdot\left[\mathbf{F}_{\alpha}^{(1)}
\times\hat{\mathbf{z}}\right]\right\}+\nabla_{\perp}\cdot\mathbf{J}_{\alpha}^{(1)}=0,
\label{n1-full}
\end{gather}
In Eq.~(\ref{n1-full}) we substitute the zero approximation
(\ref{n0}) in the term with $\Delta n_{\alpha}^{(1)}$ responsible
for weak thermal dispersion, and then, multiplying
Eq.~(\ref{n1-full}) by $4\pi e_{\alpha}$, we use
Eq.~(\ref{Pois1}). As a result, taking into account that
\begin{equation}
\frac{\omega_{p1}^{2}}{\Omega_{1}^{2}-\omega_{ii}^{2}}
+\frac{\omega_{p2}^{2}}{\Omega_{2}^{2}-\omega_{ii}^{2}}=0,
\end{equation}
one can finally obtain
\begin{widetext}
\begin{gather}
2i\omega_{ii}\frac{(\omega_{p1}^{2}+\omega_{p2}^{2})^{3}}{\omega_{p1}^{2}\omega_{p2}^{2}(\Omega_{1}^{2}
-\Omega_{2}^{2})^{2}} \frac{\partial}{\partial
t}\Delta_{\perp}\varphi^{(1)} +
\left[\frac{\omega_{p1}^{2}}{(\Omega_{1}^{2}-\omega_{ii}^{2})^{2}}v_{T1}^{2}
+\frac{\omega_{p2}^{2}}{(\Omega_{2}^{2}-\omega_{ii}^{2})^{2}}v_{T2}^{2}\right]\Delta^{2}_{\perp}\varphi^{(1)}
\nonumber \\
=4\pi
\sum_{\alpha=1,2}e_{\alpha}\nabla_{\perp}\cdot\left[\frac{i}{\omega_{ii}}\mathbf{J}_{\alpha}^{(1)}
+\frac{n_{0\alpha}}{(\omega_{ii}^{2}-\Omega_{\alpha}^{2})}
\left(\mathbf{F}_{\alpha}^{(1)}+i\frac{\Omega_{\alpha}}{\omega_{ii}}[\mathbf{F}_{\alpha}^{(1)}
\times\hat{\mathbf{z}}]\right)\right], \label{main}
\end{gather}
\end{widetext}
where the right hand side of Eq. (\ref{main}) corresponds to the
nonlinear terms. Equation (\ref{main}) can be rewritten in the
form
\begin{gather}
\frac{2i}{\omega_{ii}}\frac{\partial}{\partial
t}\Delta_{\perp}\varphi^{(1)} +
R^{2}\Delta^{2}_{\perp}\varphi^{(1)}
=\frac{4\pi\omega^{2}_{p1}\omega^{2}_{p2}(\Omega^{2}_{1}-\Omega^{2}_{2})^{2}}{\omega_{ii}^{2}(\omega_{p1}^{2}
+\omega_{p2}^{2})^{3}}
\nonumber \\
\times\sum_{\alpha=1,2}e_{\alpha}\nabla_{\perp}\cdot\left[\frac{i}{\omega_{ii}}\mathbf{J}_{\alpha}^{(1)}\right.
\nonumber \\
\left. +\frac{n_{0\alpha}}{(\omega_{ii}^{2}-\Omega_{\alpha}^{2})}
\left(\mathbf{F}_{\alpha}^{(1)}+i\frac{\Omega_{\alpha}}{\omega_{ii}}[\mathbf{F}_{\alpha}^{(1)}
\times\hat{\mathbf{z}}]\right)\right], \label{main1}
\end{gather}
where $R$ is the dispersion length defined by
\begin{equation}
\label{R} R^{2}=\frac{\omega^{2}_{p2} v^{2}_{T1}+\omega^{2}_{p1}
v^{2}_{T2}}{\omega^{2}_{p1}\Omega^{2}_{2}+\omega^{2}_{p2}\Omega^{2}_{1}}.
\end{equation}
In the linear approximation, taking $\varphi^{(1)}\sim \exp
(i\mathbf{k}_{\perp}\cdot\mathbf{r}-i\omega t)$, where $\omega$
and $\mathbf{k}_{\perp}$ the frequency and perpendicular wave
vector respectively,  Eq. (\ref{main1}) yields the dispersion
relation of ion-ion hybrid wave,
\begin{equation}
\label{disp-ion-hybr}
\omega(\mathbf{k})=\omega_{ii}\left(1+\frac{k^{2}_{\perp}R^{2}}{2}\right).
\end{equation}
From Eqs.  (\ref{motion_eq}) and (\ref{continuity_eq}), for the
second harmonic perturbations $\mathbf{v}_{\alpha}^{(2)}$ and
$n_{\alpha}^{(2)}$ we have,
\begin{equation}
\label{v2}
\mathbf{v}_{\alpha}^{(2)}=\frac{2ie_{\alpha}\omega_{ii}}{m_{\alpha}(\Omega_{\alpha}^{2}-4\omega_{ii}^{2})}
\left(\nabla_{\perp}\varphi^{(2)}+i\frac{\Omega_{\alpha}}{2\omega_{ii}}[\nabla_{\perp}\varphi^{(2)}
\times\hat{\mathbf{z}}]\right),
\end{equation}
and
\begin{equation}
\label{n2}
n_{\alpha}^{(2)}=\frac{1}{2i\omega_{ii}}\left[n_{0\alpha}\nabla_{\perp}\cdot\mathbf{v}_{\alpha}^{(2)}
+\nabla_{\perp}\cdot
(n_{0\alpha}^{(1)}\mathbf{v}_{0\alpha}^{(1)})\right],
\end{equation}
respectively, where $\mathbf{v}_{0\alpha}^{(1)}$ and
$n_{0\alpha}^{(1)}$ are determined by Eqs. (\ref{v0}) and
(\ref{n0}). The perturbation of the electrostatic potential at the
second harmonics $\varphi^{(2)}$ is determined from Eq.
(\ref{Poisson}),
\begin{equation}
\label{poisson2} \Delta_{\perp}\varphi^{(2)}=-4\pi
(e_{1}n_{1}^{(2)}+e_{2}n_{2}^{(2)}),
\end{equation}
where we neglect the electron contribution at the frequency
$2\omega_{ii}$ as before at $\omega_{ii}$.

\subsection{low-frequency disturbances}

For the LF disturbances with $\omega\ll \Omega_{1},\Omega_{2}$,
ions of both species are strongly magnetized and move only along
the external magnetic field. Then, the LF motion is governed by
the continuity equation,
\begin{equation}
\label{ns} \frac{\partial \delta n_{\alpha}}{\partial
t}+n_{0\alpha}\frac{\partial v_{\alpha,z}^{(s)}}{\partial z}=0,
\end{equation}
and the parallel momentum equation for each ion species
$\alpha=1,2$,
\begin{equation}
\label{vsi} \frac{\partial v^{(s)}_{\alpha,z}}{\partial
t}+F_{\alpha,z}=-\frac{e_{\alpha}}{m_{\alpha}}\frac{\partial\varphi^{(s)}}{\partial
z} -\frac{v_{T\alpha}^{2}}{n_{0\alpha}}\frac{\partial \delta
n_{\alpha}}{\partial z},
\end{equation}
where
\begin{equation}
\label{Fzi} F_{\alpha,z}=\langle
(\mathbf{v}_{\alpha,\perp}\cdot\nabla_{\perp})v_{\alpha,z}\rangle
\end{equation}
is the ponderomotive force (per unit ion mass) acting on the ions
due to the high-frequency (HF) pressure of the ion-ion hybrid
waves, and the angular brackets denote the average over the fast
time. From Eqs. (\ref{ns}) and (\ref{vsi}) we have,
\begin{equation}
\label{LF-ions} \frac{\partial^{2} \delta n_{\alpha}}{\partial
t^{2}}-v_{T\alpha}^{2} \frac{\partial^{2} \delta
n_{\alpha}}{\partial
z^{2}}-\frac{e_{\alpha}n_{0\alpha}}{m_{\alpha}}\frac{\partial^{2}
\varphi^{(s)}}{\partial z^{2}}-n_{0\alpha}\frac{\partial
F_{\alpha,z}}{\partial z}=0.
\end{equation}
For inertialess electrons in slow motions, one can write the force
balance equation along the magnetic field,
\begin{equation}
\label{vse}
F_{e,z}=\frac{e}{m_{e}}\frac{\partial\varphi^{(s)}}{\partial z}
-\frac{v_{Te}^{2}}{n_{0}}\frac{\partial \delta n_{e}}{\partial z},
\end{equation}
 where
\begin{equation}
\label{Fz} F_{e,z}=\langle
(\mathbf{v}_{e,\perp}\cdot\nabla_{\perp})v_{e,z}\rangle
\end{equation}
is the ponderomotive force (per unit electron mass) acting on the
electrons. From Eqs. (\ref{v}), (\ref{Fzi}) and (\ref{Fz}) we
find,
\begin{gather}
F_{\alpha,z}=\frac{1}{4}\left[(\mathbf{v}_{\alpha,\perp}^{(1)\ast}\cdot\nabla_{\perp})v_{\alpha,z}^{(1)}
+(\mathbf{v}_{\alpha,\perp}^{(1)}\cdot\nabla)v_{\alpha,z}^{(1)\ast}\right],
\label{Fzi1}
\\
F_{e,z}=\frac{1}{4}\left[(\mathbf{v}_{e,\perp}^{(1)\ast}\cdot\nabla_{\perp})v_{e,z}^{(1)}
+(\mathbf{v}_{e,\perp}^{(1)}\cdot\nabla)v_{e,z}^{(1)\ast}\right].
\label{Fz1}
\end{gather}
The expressions for the perpendicular ion
$\mathbf{v}_{\alpha,\perp}^{(1)}$ and electron
$\mathbf{v}_{e,\perp}^{(1)}$ velocities follow from Eq. (\ref{v0})
and are given by
\begin{equation}
\label{vperpi}
\mathbf{v}_{\alpha,\perp}^{(1)}=\frac{ie_{\alpha}\omega_{ii}}{m_{\alpha}(\Omega_{\alpha}^{2}-\omega_{ii}^{2})}
\left(\nabla_{\perp}\varphi^{(1)}+i\frac{\Omega_{\alpha}}{\omega_{ii}}[\nabla_{\perp}\varphi^{(1)}
\times\hat{\mathbf{z}}]\right),
\end{equation}
and
\begin{equation}
\label{vperp}
\mathbf{v}_{e,\perp}^{(1)}=-\frac{ie\omega_{ii}}{m_{e}\Omega_{e}^{2}}
\left(\nabla_{\perp}\varphi^{(1)}+i\frac{\Omega_{e}}{\omega_{ii}}[\nabla_{\perp}\varphi^{(1)}
\times\hat{\mathbf{z}}]\right),
\end{equation}
respectively. For parallel ion $v_{\alpha,z}^{(1)}$ and electron
$v_{e,z}^{(1)}$ velocities, from Eqs. (\ref{motion_eq}) and
(\ref{v}) we have
\begin{gather}
v_{\alpha,z}^{(1)}=-\frac{ie_{\alpha}}{m_{\alpha}\omega_{ii}}\frac{\partial
\varphi^{(1)}}{\partial z}, \label{vzi}
\\
v_{e,z}^{(1)}=\frac{ie}{m_{e}\omega_{ii}}\frac{\partial
\varphi^{(1)}}{\partial z}. \label{vz}
\end{gather}
Inserting Eqs. (\ref{vperpi}) and (\ref{vzi}) into Eq.
(\ref{Fzi1}) yields
\begin{equation}
\label{Fz2i}
F_{\alpha,z}=\frac{e^{2}_{\alpha}}{4m_{\alpha}^{2}(\Omega_{\alpha}^{2}-\omega_{ii}^{2})}
\frac{\partial}{\partial
z}\left(i\frac{\Omega_{\alpha}}{\omega_{ii}}\{\varphi^{(1)},\varphi^{(1)\ast}\}
-|\nabla_{\perp}\varphi^{(1)}|^{2}\right),
\end{equation}
and inserting Eqs. (\ref{vperp}) and (\ref{vz}) into Eq.
(\ref{Fz1}),
\begin{equation}
\label{Fz2} F_{e,z}=\frac{e^{2}}{4m_{e}^{2}\Omega_{e}^{2}}
\frac{\partial}{\partial
z}\left(i\frac{\Omega_{e}}{\omega_{ii}}\{\varphi^{(1)},\varphi^{(1)\ast}\}
-|\nabla_{\perp}\varphi^{(1)}|^{2}\right),
\end{equation}
where we have introduced the notation for the Poisson bracket
(Jacobian)
\begin{equation}
\label{bracket} \{f,g\}=\frac{\partial f}{\partial
x}\frac{\partial g}{\partial y}-\frac{\partial f}{\partial
y}\frac{\partial g}{\partial x}\equiv [\nabla_{\perp} f\times
\nabla_{\perp} g]\cdot\hat{\mathbf{z}}.
\end{equation}
Note that $i\{\varphi^{(1)},\varphi^{(1)\ast}\}$ is real. The
first and second terms in brackets of Eqs. (\ref{Fz2i}) and
(\ref{Fz2}) correspond to vector and scalar nonlinearities. From
Eqs. (\ref{vse}) and (\ref{Fz2}) we can evaluate the LF potential
$\varphi^{(s)}$,
\begin{equation}
\label{LF-potential}
\varphi^{(s)}=\frac{ie}{4m_{e}\Omega_{e}\omega_{ii}}\{\varphi^{(1)},\varphi^{(1)\ast}\}
+\frac{T_{e}}{e}\frac{\delta n_{e}}{n_{0}}.
\end{equation}
The quasineutrality condition reads
\begin{equation}
\label{quasineutral} e_{1}\delta n_{1}+e_{2}\delta n_{2}-e\delta
n_{e}=0.
\end{equation}
Then substitution Eq. (\ref{LF-potential}) into Eq.
(\ref{LF-ions}), taking into account Eqs. (\ref{Fz2i}) and
(\ref{quasineutral}), gives after some transformations,
\begin{gather}
\frac{\partial^{2}\delta n_{1}}{\partial
t^{2}}-(v_{T1}^{2}+\nu_{1}Z_{1}^{2}v_{s1}^{2})\frac{\partial^{2}\delta
n_{1}}{\partial
z^{2}}-\nu_{1}Z_{1}Z_{2}v_{s1}^{2}\frac{\partial^{2}\delta
n_{2}}{\partial z^{2}}
\nonumber \\
=\frac{(\omega_{p1}^{2}+\omega_{p2}^{2})}{16\pi
m_{1}(\Omega_{2}^{2}-\Omega_{1}^{2})}\frac{\partial^{2}}{\partial
z^{2}}\left[|\nabla_{\perp}\varphi^{(1)}|^{2} \right.
\nonumber \\
\left.
+i\frac{(\omega_{ii}^{2}-2\Omega_{1}^{2})}{\omega_{ii}\Omega_{1}}\{\varphi^{(1)},\varphi^{(1)\ast}\}\right],
 \label{n1s}
\end{gather}
and
\begin{gather}
\frac{\partial^{2}\delta n_{2}}{\partial
t^{2}}-(v_{T2}^{2}+\nu_{2}Z_{2}^{2}v_{s2}^{2})\frac{\partial^{2}\delta
n_{2}}{\partial
z^{2}}-\nu_{2}Z_{1}Z_{2}v_{s2}^{2}\frac{\partial^{2}\delta
n_{1}}{\partial z^{2}}
\nonumber \\
=\frac{(\omega_{p1}^{2}+\omega_{p2}^{2})}{16\pi
m_{2}(\Omega_{1}^{2}-\Omega_{2}^{2})}\frac{\partial^{2}}{\partial
z^{2}}\left[|\nabla_{\perp}\varphi^{(1)}|^{2} \right.
\nonumber \\
\left.
+i\frac{(\omega_{ii}^{2}-2\Omega_{2}^{2})}{\omega_{ii}\Omega_{2}}\{\varphi^{(1)},\varphi^{(1)\ast}\}\right],
 \label{n2s}
\end{gather}
where $v_{s1}=\sqrt{T_{e}/m_{1}}$ and $v_{s2}=\sqrt{T_{e}/m_{2}}$
are the ion sound speeds of species 1 and 2, respectively, and we
have introduced the notation for relative ion concentration
$\nu_{\alpha}=n_{0\alpha}/n_{0}$, ($\alpha=1,2$). When obtaining
Eqs. (\ref{n1s}) and (\ref{n2s}), we took into account that the
electron contribution to the scalar nonlinearity turns out to be
smaller than the contributions of other terms by a factor of order
$\sim m_{\alpha}/m_{e}$, and thus it can be neglected. The
contributions of the electron and ion vector nonlinearities, as
well as the ion scalar nonlinearity, are of the same order. In the
linear approximation, assuming $\delta n_{\alpha}\sim \exp
(ik_{z}z-i\Omega t)$, Eqs. (\ref{n1s}) and (\ref{n2s}) give the
dispersion relation
\begin{gather}
\Omega_{\pm}^{2}(\mathbf{k})=\frac{k_{z}^{2}}{2}\left\{
\left(v^{2}_{T1}+v^{2}_{T2}+\nu_{1}Z_{1}^{2}v^{2}_{s1}+\nu_{2}Z_{2}^{2}v^{2}_{s2}\right)
\right.
 \nonumber \\
\left. \pm
\sqrt{\left(v^{2}_{T1}-v^{2}_{T2}+\nu_{1}Z_{1}^{2}v^{2}_{s1}
-\nu_{2}Z_{2}^{2}v^{2}_{s2}\right)^{2}+\nu_{1}\nu_{2}Z_{1}^{2}Z_{2}^{2}v_{s1}^{2}v_{s2}^{2}}\right\},
\label{disp-LF}
\end{gather}
where $k_{z}$ is the parallel wave number. The linear dispersion
relation (\ref{disp-LF}) corresponding to two modes in a plasma
with two ion species was first obtained in Ref.~\cite{Mann1997}.
Introducing notation $q=(\mathbf{k},\omega)$ and using the
convolution identity ($f$ and $g$ are arbitrary functions)
\begin{equation}
(fg)_{q}=\int\hat{f}_{q1}\hat{g}_{q2}\delta
(q-q_{1}-q_{2})\,d^{4}q_{1}d^{4}q_{2},
\end{equation}
from Eqs. (\ref{n1s}) and (\ref{n2s}) one can write explicit
expressions for the ion density perturbations in the Fourier space
(taking $\sim\exp (i\mathbf{k}\cdot\mathbf{r}-i\Omega t)$) as
\begin{gather}
\delta
n_{1,q}=\frac{(\omega_{p1}^{2}+\omega_{p2}^{2})k_{z}^{2}}{16\pi
(\Omega_{1}^{2}-\Omega_{2}^{2})(\Omega^{2}-\Omega_{+}^{2})(\Omega^{2}-\Omega_{-}^{2})}
\nonumber \\
\times \left\{\frac{[k_{z}^{2}(v_{T2}^{2}+\nu_{2}
Z_{2}^{2}v_{s2}^{2})-\Omega^{2}]}{m_{1}}\mathcal{N}_{1,q}+\frac{k_{z}^{2}\nu_{1}
Z_{1}Z_{2}v_{s1}^{2}}{m_{2}}\mathcal{N}_{2,q}\right\},
\label{n1-four}
\\
\delta
n_{2,q}=\frac{(\omega_{p1}^{2}+\omega_{p2}^{2})k_{z}^{2}}{16\pi
(\Omega_{2}^{2}-\Omega_{1}^{2})(\Omega^{2}-\Omega_{+}^{2})(\Omega^{2}-\Omega_{-}^{2})}
\nonumber \\
\times \left\{\frac{[k_{z}^{2}(v_{T1}^{2}+\nu_{1}
Z_{1}^{2}v_{s1}^{2})-\Omega^{2}]}{m_{2}}\mathcal{N}_{2,q}+\frac{k_{z}^{2}\nu_{2}
Z_{1}Z_{2}v_{s2}^{2}}{m_{1}}\mathcal{N}_{1,q}\right\},
\label{n2-four}
\end{gather}
where
\begin{gather}
\mathcal{N}_{1,q}=\int\left[\mathbf{k}_{\perp
1}\cdot\mathbf{k}_{\perp
2}+i\frac{(\omega_{ii}^{2}-2\Omega_{1}^{2})}{\Omega_{1}\omega_{ii}}
(\mathbf{k}_{\perp,1}\times\mathbf{k}_{\perp,2})\cdot\hat{\mathbf{z}}\right]
\nonumber
\\
\times \,\varphi_{q1}\varphi_{q2}^{\ast}\delta
(q-q_{1}-q_{2})d^{4}q,
\\
\mathcal{N}_{2,q}=\int\left[\mathbf{k}_{\perp
1}\cdot\mathbf{k}_{\perp
2}+i\frac{(\omega_{ii}^{2}-2\Omega_{2}^{2})}{\Omega_{2}\omega_{ii}}
(\mathbf{k}_{\perp,1}\times\mathbf{k}_{\perp,2})\cdot\hat{\mathbf{z}}\right]
\nonumber
\\
\times \,\varphi_{q1}\varphi_{q2}^{\ast}\delta
(q-q_{1}-q_{2})d^{4}q.
\end{gather}
Equations (\ref{n1s}) and (\ref{n2s}) describe the dynamics of LF
acoustic-type disturbances (in the linear case corresponding to
two branches of ion-ion sound) under the action of the
ponderomotive force of the HF field of ion-ion hybrid wave.

\subsection{neglecting second harmonics}

Equations (\ref{main1}), (\ref{v2})-(\ref{poisson2}) for the HF
motions along with Eqs. (\ref{n1s}) and (\ref{n2s}) for LF
disturbances is a closed system of nonlinear equations for the HF
envelope of electrostatic potential $\varphi^{(1)}$ and LF ion
density perturbations $\delta n_{1}$ and $\delta n_{2}$. It can be
seen, however, that due to taking into account the nonlinear terms
corresponding to the second harmonic of the ion-ion hybrid
frequency (that is, containing $\varphi^{(2)}$,
$\mathbf{v}_{\alpha}^{(2)}$, and $n_{\alpha}^{(2)}$), this system
turns out to be extremely cumbersome and very difficult to
analyze. In particular, taking into account second harmonics leads
to terms containing $\mathbf{F}_{\alpha}$. In addition, vector
nonlinearities have, generally speaking, the same order as scalar
ones. This distinguishes the case under consideration from the
case of nonlinear upper-hybrid waves, where, under reasonable
conditions, vector nonlinearity can be neglected, as well as from
the case of lower-hybrid waves, when, on the contrary, vector
nonlinearity is always dominant. However, the system of equations
(\ref{main1}) and (\ref{v2})-(\ref{poisson2}) is greatly
simplified for radially symmetric field distributions. We show
below in Sec. V that taking into account the contribution of the
second harmonics results in the existence of a stable 2D soliton,
but for now we neglect this contribution. Then, Eq. (\ref{main1})
takes the form (here and after denoting $\varphi=\varphi^{(1)}$)
\begin{gather}
\frac{2i}{\omega_{ii}}\frac{\partial}{\partial t}\Delta\varphi +
R^{2}\Delta^{2}\varphi =\frac{
\omega^{2}_{p1}\omega^{2}_{p2}(\Omega^{2}_{1}-\Omega^{2}_{2})^{2}}{\omega_{ii}^{2}(\omega_{p1}^{2}
+\omega_{p2}^{2})^{3}}
\nonumber \\
\times\sum_{\alpha=1,2}\frac{\omega_{p\alpha}^{2}}{
(\omega^{2}_{ii}-\Omega^{2}_{\alpha})n_{0\alpha}}\left[\nabla_{\perp}\cdot\left(\delta
n_{\alpha}\nabla_{\perp}\varphi\right)+i\frac{\Omega_{\alpha}}{\omega_{ii}}\{\delta
n_{\alpha},\varphi\}\right] , \label{main2}
\end{gather}
which can be rewritten as
\begin{gather}
\frac{2i}{\omega_{ii}}\frac{\partial}{\partial
t}\Delta_{\perp}\varphi + R^{2}\Delta^{2}_{\perp}\varphi =\frac{
\omega^{2}_{p1}\omega^{2}_{p2}(\Omega^{2}_{2}-\Omega^{2}_{1})}{\omega_{ii}^{2}(\omega_{p1}^{2}
+\omega_{p2}^{2})^{2}}
\nonumber \\
\times\nabla_{\perp}\cdot\left[\left(\frac{\delta
n_{1}}{n_{01}}-\frac{\delta
n_{2}}{n_{02}}\right)\nabla_{\perp}\varphi \right.
\nonumber \\
\left.
 +\frac{i}{\omega_{ii}}\left(\Omega_{1}\frac{\delta
n_{1}}{n_{01}}-\Omega_{2}\frac{\delta n_{2}}{n_{02}}\right)
(\nabla_{\perp}\varphi\times\hat{\mathbf{z}})\right].
\label{main3}
\end{gather}
The first term in square brackets on the right hand side of Eq.
(\ref{main2}) corresponds to the scalar nonlinearity, and the
second term in the form of the Poisson bracket corresponds to the
so-called vector nonlinearity. The latter identically vanishes in
the one-dimensional case and also for radially symmetric field
distributions.

\section{\label{non-disp-rel} nonlinear dispersion relation }

In this section we consider the linear theory of the modulational
instability of a pump wave with a frequency close to the ion-ion
hybrid frequency $\omega_{ii}$ in the framework of the model
equations (\ref{n1s}), (\ref{n2s}) and (\ref{main2}). We decompose
the ion-ion hybrid wave into the pump wave and two sidebands, i.e.
\begin{gather}
\varphi=\varphi_{0}\mathrm{e}^{i\mathbf{k}_{\perp
}\cdot\mathbf{r}_{\perp}-i\delta_{\mathbf{k}}t}
+\varphi_{+}\mathrm{e}^{i(\mathbf{k}_{\perp }
+\mathbf{q}_{\perp})\cdot\mathbf{r}_{\perp}-i(\delta_{\mathbf{k}}+\Omega)t}
\nonumber \\
+\varphi_{-}\mathrm{e}^{i(\mathbf{k}_{\perp}
-\mathbf{q}_{\perp})\cdot\mathbf{r}_{\perp}-i(\delta_{\mathbf{k}}-\Omega)t}
+ \mathrm{c.c.} ,
\end{gather}
where $\delta_{\mathbf{k}}=\omega_{ii}k_{\perp}^{2}R^{2}/2$, while
the low frequency perturbations of ion plasma densities are
expressed as
\begin{equation}
\label{n-four} \delta
n_{\alpha}=\hat{n}_{\alpha}\mathrm{e}^{i\mathbf{q}\cdot\mathbf{r}-i\Omega
t}+ \mathrm{c.c.} .
\end{equation}
The amplitudes of the up-shifted $\varphi_{+}$ and down-shifted
$\varphi_{-}$ satellites can be calculated from Eq.~(\ref{main3}).
We have
\begin{gather}
D_{+}\varphi_{+}=(A_{+,1}\hat{n}_{1}-A_{+,2}\hat{n}_{2})\varphi_{0},
\label{D-plus}
\\
D_{-}\varphi_{-}^{\ast}=(A_{-,1}\hat{n}_{1}-A_{-,2}\hat{n}_{2})\varphi_{0}^{\ast},
\label{D-minus}
\end{gather}
where  $D_{\pm}$ is the Fourier transform of the linear operator
in the left hand side of Eq. (\ref{main2}) evaluated in
$\mathbf{k}\pm\mathbf{q}$ and $\delta_{\mathbf{k}}\pm\Omega$,
\begin{gather}
D_{\pm}=\frac{2}{\omega_{ii}}(\mathbf{k}_{\perp}
\pm\mathbf{q}_{\perp})^{2}(\delta_{\pm}\mp\Omega),
\\
\delta_{\pm}=\omega_{ii}[(\mathbf{k}_{\perp}
\pm\mathbf{q}_{\perp})^{2}-\mathbf{k}_{\perp}^{2}]R^{2}/2,
\end{gather}
and
\begin{gather}
A_{\pm,\alpha}=-\frac{
\omega^{2}_{p1}\omega^{2}_{p2}(\Omega^{2}_{2}-\Omega^{2}_{1})}{\omega_{ii}^{2}(\omega_{p1}^{2}
+\omega_{p2}^{2})^{2}n_{0\alpha}}
\nonumber \\
\times\left[ (\mathbf{k}_{\perp}^{2}\pm\mathbf{k}_{\perp
}\cdot\mathbf{q}_{\perp})\pm
i\frac{\Omega_{\alpha}}{\omega_{ii}}(\mathbf{k}_{\perp
}\times\mathbf{q}_{\perp})\cdot\hat{\mathbf{z}}\right].
\end{gather}
The amplitudes of the LF perturbations $\hat{n}_{1}$ and
$\hat{n}_{2}$ can be found from Eqs. (\ref{n1-four}) and
(\ref{n2-four}),
\begin{gather}
\hat{n}_{1}=\frac{q_{z}^{2}(B_{+,1}\varphi_{0}^{\ast}\varphi_{+}
+B_{-,1}\varphi_{0}\varphi_{-}^{\ast})}{
[\Omega^{2}-\Omega_{+}^{2}(q_{z})][\Omega^{2}-\Omega_{-}^{2}(q_{z})]},
\label{n1f}\\
\hat{n}_{2}=-\frac{q_{z}^{2}(B_{+,2}\varphi_{0}^{\ast}\varphi_{+}
+B_{-,2}\varphi_{0}\varphi_{-}^{\ast})}{
[\Omega^{2}-\Omega_{+}^{2}(q_{z})][\Omega^{2}-\Omega_{-}^{2}(q_{z})]},
\label{n2f}
\end{gather}
where $\Omega_{+}^{2}$ and $\Omega_{-}^{2}$ are determined by
Eq.~(\ref{disp-LF}), and
\begin{gather}
B_{\pm,1}=\frac{q_{z}^{2}(v_{T2}^{2}+\nu_{2}Z_{2}^{2}v_{s2}^{2})-\Omega^{2}}{m_{1}}\mathcal{B}_{\pm,1}
+\frac{q_{z}^{2}\nu_{1}Z_{1}Z_{2}v_{s1}^{2}}{m_{2}}\mathcal{B}_{\pm,2},
\\
B_{\pm,2}=\frac{q_{z}^{2}(v_{T1}^{2}+\nu_{1}Z_{1}^{2}v_{s1}^{2})-\Omega^{2}}{m_{2}}\mathcal{B}_{\pm,2}
+\frac{q_{z}^{2}\nu_{2}Z_{1}Z_{2}v_{s2}^{2}}{m_{1}}\mathcal{B}_{\pm,1},
\end{gather}
where
\begin{gather}
\mathcal{B}_{\pm,\alpha}=\frac{(\omega_{p1}^{2}+\omega_{p2}^{2})}{16\pi
(\Omega_{1}^{2}-\Omega_{2}^{2})}
\left[(\mathbf{k}_{\perp}^{2}\pm\mathbf{k}_{\perp}\cdot\mathbf{q})\right.
\nonumber \\
\left. \pm i\frac{(\omega_{ii}^{2}
-2\Omega_{\alpha}^{2})}{\Omega_{\alpha}\omega_{ii}}(\mathbf{k}_{\perp}\times\mathbf{q}_{\perp})
\cdot\hat{\mathbf{z}}\right].
\end{gather}
By combining Eqs.~(\ref{D-plus}), (\ref{D-minus}), (\ref{n1f}),
and (\ref{n2f}) we obtain a nonlinear dispersion relation
\begin{gather}
[\Omega^{2}-\Omega_{+}^{2}(q_{z})][\Omega^{2}-\Omega_{-}^{2}(q_{z})]-q_{z}^{2}|\varphi_{0}|^{2}
\nonumber \\
\times\left[\frac{A_{+,1}B_{+,1}+A_{+,2}B_{+,2}}{D_{+}}
+\frac{A_{-,1}B_{-,1}+A_{-,2}B_{-,2}}{D_{-}}\right]=0.
\label{disp-non}
\end{gather}
Note, that in the case of coplanar (in the plane perpendicular to
the magnetic field) wave vectors
$\mathbf{k}_{\perp}\parallel\mathbf{q}_{\perp}$, the parametric
coupling of the waves due to the vector nonlinearity is absent,
while the coupling due to the scalar nonlinearity is the most
effective. In general case
$\mathbf{k}_{\perp}\nparallel\mathbf{q}_{\perp}$ both types of the
nonlinearities yield comparable contribution, and this, taking
into account that Eq. (\ref{disp-non}) is an equation of the sixth
degree in $\Omega$, leads to a rather complicated picture of
instability. Significant simplifications are possible in a number
of special cases. For example, assuming that $\Omega\ll
\Omega_{-}, \Omega_{+}$,
$\mathbf{k}_{\perp}\parallel\mathbf{q}_{\perp}$ and
$\mathbf{k}_{\perp}\gg \mathbf{q}_{\perp}$, the dispersion
equation (\ref{disp-non}) after direct calculations can be reduced
to the form
\begin{equation}
\label{growth1}
\frac{\omega_{ii}^{2}q_{\perp}^{4}R^{4}}{4}-(\Omega-q_{\perp}v_{g})^{2}
=\frac{|E_{0}|^{2}\omega_{p1}^{2}\omega_{p2}^{2}q_{\perp}^{2}R^{2}F}{32\pi(\omega_{p1}^{2}
+\omega_{p2}^{2})n_{0}\nu_{1}\nu_{2} G},
\end{equation}
where $v_{g}=\omega_{ii}k_{\perp} R$ is the group velocity of the
ion-ion hybrid wave, and we have introduced the notations for the
coefficients $F$ and $G$,
\begin{gather}
F=\nu_{1} T_{1}+\nu_{2} T_{2}+T_{e}(\nu_{1} Z_{1}+\nu_{2}
Z_{2})^{2},
\label{F} \\
G=T_{1}T_{2}
+\nu_{2}Z_{2}^{2}T_{e}T_{1}+\nu_{1}Z_{1}^{2}T_{e}T_{2}, \label{G}
\end{gather}
which we will use in what follows. Equation (\ref{growth1})
predicts instability with the growth rate
$\gamma=|\mathrm{Im}\,\Omega|$,
\begin{equation}
\label{grow} \gamma=\frac{\omega_{ii}q_{\perp}^{2}R^{2}}{2}
 \left[\frac{|E_{0}|^{2}\omega_{p1}^{2}\omega_{p2}^{2}F}{8\pi\omega_{ii}^{2}(\omega_{p1}^{2}
+\omega_{p2}^{2})q_{\perp}^{2}R^{2}n_{0}\nu_{1}\nu_{2}
G}-1\right]^{1/2}.
\end{equation}
In the opposite case $\mathbf{k}_{\perp}\ll \mathbf{q}_{\perp}$,
one can get
\begin{equation}
\label{growth2}
\frac{\omega_{ii}^{2}q_{\perp}^{4}R^{4}}{4}-\Omega^{2}
=\frac{|E_{0}|^{2}\omega_{p1}^{2}\omega_{p2}^{2}q_{\perp}^{2}R^{2}F}{32\pi(\omega_{p1}^{2}
+\omega_{p2}^{2})n_{0}\nu_{1}\nu_{2} G},
\end{equation}
and Eq. (\ref{growth2}) describes a purely growing instability
with the growth rate given by Eq. (\ref{grow}). For example, for
the upper $F$ region/topside ionosphere ($\sim 500$ km), taking
$n_{0}\sim 5\cdot 10^{11}$ $\mathrm{m}^{-3}$, $n_{1}/n_{2}\sim
10$, $T_{1,2}\sim 2\cdot 10^{3}$ $\mathrm{K}$, and $\Omega_{1}\sim
2\cdot 10^{2}$ $\mathrm{s}^{-1}$ in accordance with
Ref.~\cite{Horvitz1990} (subscripts 1 and 2 correspond to
$\mathrm{O}^{+}$ and $\mathrm{He}^{+}$, respectively, and the
concentration of $\mathrm{H}^{+}$ is two orders of magnitude less
than the concentration of $\mathrm{O}^{+}$), the estimate for the
threshold field at $q_{\perp}R\sim 0.1$ is $E_{0}\sim 100$
$\mathrm{mV}/\mathrm{m}$.

\section{collapse of ion-ion hybrid waves}

In this section, neglecting the second harmonics, we discuss the
possibility of collapse of the ion-ion hybrid waves. Let us
consider an important case when we can neglect the time
derivatives in the LF equations (\ref{n1s}) and (\ref{n2s}).
Physically, this corresponds to the balance between gas-kinetic
and wave (ponderomotive) pressures. In this static approximation
("subsonic" case), from Eqs. (\ref{n1s}) and (\ref{n2s}) for LF
perturbations of ion densities $\delta n_{1}$ and $\delta n_{2}$
one can obtain,
\begin{gather}
\delta
n_{1}=\frac{(\omega_{p1}^{2}+\omega_{p2}^{2})[(T_{2}+\nu_{2}Z_{2}^{2}T_{e})
\mathcal{N}_{1}+\nu_{1}Z_{1}Z_{2}T_{e}\mathcal{N}_{2}]}{16\pi
(\Omega_{1}^{2}-\Omega_{2}^{2})G},
\label{n1-subson} \\
\delta n_{2}=\frac{(\omega_{p1}^{2}+\omega_{p2}^{2})
[(T_{1}+\nu_{1}Z_{1}^{2}T_{e})\mathcal{N}_{2}
 +\nu_{2}Z_{1}Z_{2}T_{e}\mathcal{N}_{1}]}{16\pi
(\Omega_{2}^{2}-\Omega_{1}^{2})G}, \label{n2-subson}
\end{gather}
where
\begin{gather}
\mathcal{N}_{1}= \left[|\nabla_{\perp}\varphi^{(1)}|^{2}
+i\frac{(\omega_{ii}^{2}-2\Omega_{1}^{2})}{\omega_{ii}\Omega_{1}}\{\varphi^{(1)},
\varphi^{(1)\ast}\}\right],
 \label{NN1-subson}
 \\
\mathcal{N}_{2}=\left[|\nabla_{\perp}\varphi^{(1)}|^{2}
+i\frac{(\omega_{ii}^{2}-2\Omega_{2}^{2})}{\omega_{ii}\Omega_{2}}\{\varphi^{(1)},
\varphi^{(1)\ast}\}\right],
 \label{NN2-subson}
\end{gather}
and $G$ is determined by Eq. (\ref{G}). The perturbation of ion
densities due to the vector nonlinearity can correspond to both a
density well and a hump and depends on the relative phase of
$\varphi$ and $\varphi^{\ast}$. Introducing dimensionless
variables by
\begin{gather}
t\rightarrow \frac{\omega_{ii}t}{2}, \quad \mathbf{r}_{\perp}
\rightarrow \frac{\mathbf{r}_{\perp}}{R}, \label{dimen1}
\\
 \psi \rightarrow
\frac{\omega_{p1}\omega_{p2}}{4\omega_{ii}R}\left[\frac{F}{\pi
n_{0}\nu_{1}\nu_{2}(\omega_{p1}^{2}+\omega_{p2}^{2})G}\right]^{1/2}\varphi,
\label{dimen2}
\end{gather}
where $F$ is determined by Eq. (\ref{F}), and substituting Eqs.
(\ref{n1-subson}) and (\ref{n2-subson}) into Eq. (\ref{main3}),
one can obtain
\begin{gather}
i\frac{\partial}{\partial t}\Delta_{\perp}\psi +
\Delta^{2}_{\perp}\psi +\nabla_{\perp}\cdot
(|\nabla_{\perp}\psi|^{2}\nabla_{\perp}\psi)+c_{1}\{\{\psi,\psi^{\ast}\},\psi\}
\nonumber \\
+ic_{2}\{|\nabla_{\perp}\psi|^{2},\psi\}+ic_{3}\nabla_{\perp}\cdot(\{\psi,\psi^{\ast}\}\nabla_{\perp}\psi)=0,
\label{main4}
\end{gather}
where we have introduced the dimensionless coefficients $c_{1}$,
$c_{2}$ and $c_{3}$,
\begin{equation}
\label{G-coef} c_{1}=\frac{G_{1}}{F},\quad c_{2}=\frac{G_{2}}{F},
\quad c_{3}=\frac{G_{3}}{F},
\end{equation}
where $G_{1}$, $G_{2}$ and $G_{3}$ are determined in the Appendix.
The nonlinear equation (\ref{main4}) for the envelope $\psi$
differs significantly from the corresponding equations (in the
subsonic case) for nonlinear Langmuir, lower-hybrid and
upper-hybrid waves. If there is only scalar nonlinearity, that is
$c_{1}=0$, $c_{2}=0$ and $c_{3}=0$, Eq. (\ref{main4}) is reduced
to the well-known equation for nonlinear Langmuir waves
\cite{Zakharov1972}. The case $c_{2}=0$ and $c_{3}=0$ corresponds
to nonlinear upper-hybrid waves (with another dimensionless
coefficient $c_{1}$) \cite{Lashkin2007}. If the scalar
nonlinearity can be neglected, and also if $c_{2}=0$ and $c_{3}=0$
(again with another dimensionless coefficient $c_{1}$), Eq.
(\ref{main4}) becomes the equation for the envelope at the
lower-hybrid frequency
\cite{Sturman1976,Shapiro1984,Shapiro1993,Shapiro1995}. In all
these cases, the corresponding dimensionless variables for time
and space coordinates and the electrostatic potential envelope are
implied. The nonlinearities corresponding to the fifth ($c_{2}\neq
0$) and sixth ($c_{3}\neq 0$) terms in Eq. (\ref{main4}) have
apparently never been considered in nonlinear problems before.
Equation (\ref{main4}) conserves the plasmon number
\begin{equation}
\label{Energy} N=\int |\nabla_{\perp}\psi|^{2}d^{2}\mathbf{r},
\end{equation}
the momentum
\begin{equation}
\label{P} \mathbf{P}=\int \mathbf{p}d^{2}\mathbf{r},
\end{equation}
where the momentum density is
\begin{equation}
p_{l}=\frac{i}{2}\int
(\nabla_{k}\psi^{\ast}\nabla_{l}\nabla_{k}\psi-\mathrm{c.
c.})d^{2}\mathbf{r},
\end{equation}
the angular momentum
\begin{equation}
\label{M} \mathbf{M}=\int
([\mathbf{r}\times\mathbf{p}])d^{2}\mathbf{r},
\end{equation}
and the Hamiltonian
\begin{gather}
H=\int\left[|\Delta_{\perp}\psi|^{2}-\frac{|\nabla_{\perp}\psi|^{4}}{2}
+\frac{c_{1}\{\psi,\psi^{\ast}\}^{2}}{2} \right.
\nonumber \\
\left.
-\frac{i(c_{3}-c_{2})}{2}\{\psi,\psi^{\ast}\}|\nabla_{\perp}\psi|^{2}
\right]d^{2}\mathbf{r}. \label{Hamil}
\end{gather}
Note that the second term in $H$ is a focusing nonlinearity, while
the sign of the other terms (which, as you can easily see, is real
since $\{\psi,\psi^{\ast}\}$ is purely imaginary) depend on the
phase relation between $\psi$ and $\psi^{\ast}$ and also on the
signs of the coefficients $c_{1}$, $c_{2}$ and $c_{3}$, that is,
on the ratio of ion and electron temperatures, masses of ions of
different species and their relative concentrations. An essential
feature of the considered model equation (\ref{main4}) is its
two-dimensional nature and the cubic nonlinearity. The stationary
solution of Eq. (\ref{main4}) in the form of $\psi
(\mathbf{r},t)=\Psi (\mathbf{r})\exp (i\lambda^{2}t)$ corresponds
to a stationary point $H$ for a fixed plasmon number $N$ and
resolves the variational problem
\begin{equation}
\label{var-prob} \delta (H+\lambda^{2}N)=0,
\end{equation}
that is
\begin{gather}
-\lambda^{2}\Delta_{\perp}\Psi+\Delta_{\perp}^{2}\Psi+
\nabla_{\perp}\cdot (|\nabla_{\perp}\Psi|^{2}\nabla_{\perp}\Psi)
+c_{1}\{\{\Psi,\Psi^{\ast}\},\Psi\}
\nonumber \\
+ic_{2}\{|\nabla_{\perp}\Psi|^{2},\Psi\}
+ic_{3}\nabla_{\perp}\cdot(\{\Psi,\Psi^{\ast}\}\nabla_{\perp}\Psi)=0.
\label{main5}
\end{gather}
By analogy with Langmuir, upper-hybrid \cite{Kuznetsov1986} and
lower-hybrid \cite{{KuznetsovScorich1988},Scorich2010} waves,
multiplying Eq. (\ref{main5}) by $\Psi^{\ast}$, and then
integrating over the whole $D$-dimension space, we get
\begin{equation}
\label{var1} \lambda^{2}N+I_{1}-I_{2}+I_{3}-I_{4}=0,
\end{equation}
where
\begin{gather}
I_{1}=\int |\Delta_{\perp}\Psi|^{2}d^{2}\mathbf{r}, \quad
I_{2}=\int |\nabla_{\perp}\Psi|^{4} d^{2}\mathbf{r},
\\
I_{3}=c_{1}\int \{\Psi,\Psi^{\ast}\}^{2} d^{2}\mathbf{r},
\\
I_{4}=i(c_{3}-c_{2})\int
\{\Psi,\Psi^{\ast}\}|\nabla_{\perp}\Psi|^{2} d^{2}\mathbf{r}.
\end{gather}
On the other hand, one can write
\begin{equation}
\label{var2}
H=I_{1}-\frac{I_{2}}{2}+\frac{I_{3}}{2}-\frac{I_{4}}{2}.
\end{equation}
Next, we consider an $N$-preserving scaling transformation
$\Psi^{(\alpha)}=\Psi (\alpha\mathbf{r})$ and introduce the
corresponding values
\begin{gather}
N^{(\alpha)}=\alpha^{2-D}N, \quad
I_{1}^{(\alpha)}=\alpha^{4-D}I_{1},
\\
I_{2}^{(\alpha)}=\alpha^{4-D}I_{2}, \quad
I_{3}^{(\alpha)}=\alpha^{4-D}I_{3}, \quad
I_{4}^{(\alpha)}=\alpha^{4-D}I_{4}.
\end{gather}
It is evident that
\begin{equation}
\frac{\partial}{\partial
\alpha}(H^{(\alpha)}+\lambda^{2}N^{(\alpha)})|_{\alpha=1}=0,
\end{equation}
from which we have,
\begin{gather}
\lambda^{2}(2-D)N+(4-D)I_{1}-\frac{4-D}{2}I_{2} \nonumber
\\
+\frac{4-D}{2}I_{3}-\frac{4-D}{2}I_{4}=0. \label{var3}
\end{gather}
From Eqs. (\ref{var1}), (\ref{var2}) and (\ref{var3}) we then find
\begin{equation}
\label{H-N} H=\lambda^{2}\frac{(D-2)}{(4-D)}N.
\end{equation}
Equation (\ref{H-N}) is identical to the relationship between the
Hamiltonian $H$ and the number of plasmons $N$ for nonlinear
Langmuir waves \cite{Scorich2010,Kuznetsov1986}. It can be seen
that the reason for the coincidence is the same linear parts
(dimensionless) and the cubic nature of the nonlinear terms. Since
for the considered 2D model we have $H=0$ for stationary
solutions, one can conclude that an arbitrary initial localized
field distribution with $H\neq 0$ never reaches a stationary state
in the course of evolution, that is, either spreads out or
collapses. Hamiltonian (\ref{Hamil}) is not positively definite,
despite the fact that the third term in Eq. (\ref{Hamil}), as
mentioned above, may  have a defocusing character. A rigorous
proof of the collapse of ion-ion hybrid waves (as well as Langmuir
waves in arbitrary geometry) is apparently a very difficult
problem. Here we only point out, taking into account the arguments
presented above, that with a negative initial Hamiltonian, the
collapse of two-dimensional ion-ion hybrid waves apparently
occurs.

\section{\label{sec5}
 stable 2D radially symmetric soliton}

In the radially symmetric case, the vector nonlinearities vanish
identically, and  Eqs. (\ref{v0}), (\ref{n0}) and (\ref{v2}) takes
the form
\begin{gather}
\mathbf{v}_{0\alpha}^{(1)}=\frac{ie_{\alpha}\omega_{ii}}{m_{\alpha}(\omega_{ii}^{2}-\Omega_{\alpha}^{2})}
\mathbf{E},
 \label{v000}
\\
n_{0\alpha}^{(1)}=\frac{1}{4\pi
e_{\alpha}}\frac{\omega_{p\alpha}^{2}}{(\omega_{ii}^{2}-\Omega_{\alpha}^{2})}\nabla_{\perp}\cdot\mathbf{E},
\label{n000}
\\
\mathbf{v}_{\alpha}^{(2)}=\frac{2ie_{\alpha}\omega_{ii}}{m_{\alpha}(4\omega_{ii}^{2}
-\Omega_{\alpha}^{2})}\mathbf{E}^{(2)}, \label{v22}
\end{gather}
where $\mathbf{E}=-\nabla_{\perp}\varphi^{(1)}$ and
$\mathbf{E}^{(2)}=-\nabla_{\perp}\varphi^{(2)}$ is the electric
field at the second harmonics. Equation (\ref{poisson2}) can be
rewritten as
\begin{equation}
\label{poisson3} \nabla_{\perp}\cdot\mathbf{E}^{(2)}=4\pi
\sum_{\alpha=1,2}e_{\alpha}n_{\alpha}^{(2)}
\end{equation}
From Eqs. (\ref{n2}) and (\ref{poisson3}), we have
\begin{equation}
\label{E22}
\mathbf{E}^{(2)}=\frac{2\pi}{i\omega_{ii}}\sum_{\alpha=1,2}(e_{\alpha}n_{0\alpha}
\mathbf{v}_{\alpha}^{(2)}+e_{\alpha}n_{0\alpha}^{(1)}
\mathbf{v}_{0\alpha}^{(1)}).
\end{equation}
Using Eq. (\ref{v22}), we eliminate $\mathbf{v}_{\alpha}^{(2)}$ in
Eq. (\ref{E22}) and then substitute expressions for
$n_{0\alpha}^{(1)}$ and $\mathbf{v}_{0\alpha}^{(1)}$. As a result
one can find,
\begin{equation}
\label{E222}
\mathbf{E}^{(2)}=-\frac{(\omega_{p1}^{2}+\omega_{p2}^{2})^{2}}
{8\pi\varepsilon(2\omega_{ii})(\Omega_{1}^{2}-\Omega_{2}^{2})^{2}}
\left(\frac{1}{e_{1}n_{01}}+\frac{1}{e_{2}n_{02}}\right)\mathbf{E}\nabla_{\perp}\cdot\mathbf{E},
\end{equation}
where
\begin{equation}
\varepsilon(2\omega_{ii})=1-\frac{\omega_{p1}^{2}}{4\omega_{ii}^{2}
-\Omega_{1}^{2}}-\frac{\omega_{p2}^{2}}{4\omega_{ii}^{2}
-\Omega_{2}^{2}}.
\end{equation}
Inserting Eq. (\ref{E222}) into Eq. (\ref{v22}) we have
\begin{gather}
\mathbf{v}_{\alpha}^{(2)}=\frac{ie_{\alpha}\omega_{ii}(\omega_{p1}^{2}+\omega_{p2}^{2})^{2}}
{4\pi
m_{\alpha}\varepsilon(2\omega_{ii})(\Omega_{\alpha}^{2}-4\omega_{ii}^{2})(\Omega_{1}^{2}-\Omega_{2}^{2})^{2}}
\nonumber
\\
\times\left(\frac{1}{e_{1}n_{01}}+\frac{1}{e_{2}n_{02}}\right)\mathbf{E}\nabla_{\perp}\cdot\mathbf{E}.
\label{v2222}
\end{gather}
It can be shown that the second term in Eq. (\ref{n2}) can be
neglected, and then substituting Eq. (\ref{v2222}) into Eq.
(\ref{n2})  one can obtain
\begin{gather}
n_{\alpha}^{(2)}=\frac{e_{\alpha}n_{0\alpha}(\omega_{p1}^{2}+\omega_{p2}^{2})^{2}}{8\pi
m_{\alpha}\varepsilon(2\omega_{ii})(\Omega_{1}^{2}-\Omega_{2}^{2})^{2}(\Omega_{\alpha}^{2}-4\omega_{ii}^{2})}
\nonumber
\\
\times \left(\frac{1}{e_{1}n_{01}}+\frac{1}{e_{2}n_{02}}\right)
\nabla_{\perp}\cdot(\mathbf{E}\nabla_{\perp}\cdot\mathbf{E}).
\label{n2222}
\end{gather}
The term corresponding to the contribution of the second harmonics
on the right hand side of Eq. (\ref{main1}) can be written as
$\nabla_{\perp}\cdot\mathbf{N}^{(2)}$, where
\begin{gather}
\mathbf{N}^{(2)}=\frac{4\pi\omega^{2}_{p1}\omega^{2}_{p2}(\Omega^{2}_{1}-\Omega^{2}_{2})^{2}}{\omega_{ii}^{2}(\omega_{p1}^{2}
+\omega_{p2}^{2})^{3}}\sum_{\alpha=1,2}e_{\alpha}\left\{\frac{i}{\omega_{ii}}
(n_{0\alpha}^{(1)\ast}\mathbf{v}_{\alpha}^{(2)} \right.
\nonumber \\
\left. +n_{\alpha}^{(2)}\mathbf{v}_{0\alpha}^{(1)\ast})
\!+\!\frac{n_{0\alpha}}{(\omega^{2}_{ii}-\Omega^{2}_{\alpha})}
[(\mathbf{v}_{0\alpha}^{(1)\ast}\cdot\nabla)\mathbf{v}_{\alpha}^{(2)}
\!+\!(\mathbf{v}_{\alpha}^{(2)}\cdot\nabla)\mathbf{v}_{0\alpha}^{(1)\ast}]\right\}.
\label{N2}
\end{gather}
In the considered radially symmetric case, we are interested here
only in the radial component $E_{r}$ of the electric field
$\mathbf{E}$. Then, writing Eq. (\ref{main3}) through  the
electric field $\mathbf{E}$ with the additional term which takes
into account the contribution of second harmonics,  and taking its
radial projection, we have for $E_{r}$,
\begin{gather}
\frac{2i}{\omega_{ii}}\frac{\partial E_{r}}{\partial
t}+R^{2}\left(\Delta_{r} E_{r}-\frac{E_{r}}{r^{2}}\right)=\frac{
\omega^{2}_{p1}\omega^{2}_{p2}(\Omega^{2}_{2}-\Omega^{2}_{1})}{\omega_{ii}^{2}(\omega_{p1}^{2}
+\omega_{p2}^{2})^{2}} \nonumber \\ \times \left(\frac{\delta
n_{1}}{n_{01}}-\frac{\delta n_{2}}{n_{02}}\right)
E_{r}+(\mathbf{N}^{(2)})_{r}, \label{EE-eq0}
\end{gather}
where $\Delta_{r}=\partial^{2}/\partial
r^{2}+(1/r)\partial/\partial r$ is the 2D radial Laplacian, and
for the considered 2D case we have used the relation $(\Delta
\mathbf{E})_{r}=\Delta_{r}E_{r}-E_{r}/r^{2}$. Further, as in the
previous section, we consider the static approximation
(\ref{n1-subson}) and (\ref{n2-subson}) for ion density
perturbations, and take into account that, in the radially
symmetric case, the vector nonlinearities in Eqs.
(\ref{NN1-subson}) and (\ref{NN2-subson}) vanish identically.
Next, we use the dimensionless variables defined by Eq.
(\ref{dimen1}), and introduce the dimensionless radial electric
field $E$ through
\begin{equation}
E \rightarrow
\frac{\omega_{p1}\omega_{p2}}{4\omega_{ii}}\left[\frac{F}{\pi
n_{0}\nu_{1}\nu_{2}(\omega_{p1}^{2}+\omega_{p2}^{2})G}\right]^{1/2}E_{r}.
\end{equation}
From Eqs. (\ref{n1-subson}), (\ref{n2-subson}), (\ref{N2}) and
(\ref{EE-eq0}), one can finally obtain
\begin{equation}
\label{E-eq} i\frac{\partial E}{\partial
t}+\Delta_{r}E-\frac{E}{r^{2}}+|E|^{2}E+QE\Delta_{r}|E|^{2}-\frac{Q|E|^{2}E}{r^{2}}=0,
\end{equation}
where $Q$ is determined in the Appendix. Equation (\ref{E-eq})
conserves the plasmon number
\begin{equation}
\label{N1} \tilde{N}=\int |E|^{2} d^{2}\mathbf{r},
\end{equation}
and Hamiltonian
\begin{equation}
\label{H1} \tilde{H}=\!\!\int \!
\left\{\left|\frac{1}{r}\frac{\partial (rE)}{\partial
r}\right|^{2}\!-\!\frac{|E|^{4}}{2}\!+\!\frac{Q}{2}\left(\frac{\partial
|E|^{2}}{\partial
r}\right)^{2}\!+\!\frac{Q|E|^{4}}{2r^{2}}\right\}\!d^{2}\mathbf{r}.
\end{equation}
An equation similar to Eq. (\ref{E-eq}) was obtained in
Ref.~\cite{Davydova2005}, where the influence of electron-electron
nonlinearities on unstable two-dimensional and three-dimensional
Langmuir solitons was studied. In that work it was shown that the
effective radius $r_{eff}$, defined as
\begin{equation}
\label{r-eff}  r_{eff}^{2}=\frac{1}{\tilde{N}}\int
r^{2}|E|^{2}d^{2}\mathbf{r},
\end{equation}
is bounded from below provided $Q\neq 0$, so that additional
nonlinear terms proportional to $Q$ prevent collapse (the same
applies to the 3D case).  Moreover, it has also been shown that
the gradient norm $\int |\partial E/\partial
r|^{2}d^{2}\mathbf{r}$ is bounded from above by conserved
quantities $\tilde{N}$ and $\tilde{H}$. The authors of
Ref.~\cite{Davydova2005} also numerically found the 2D soliton
solution and demonstrated the stability of such a soliton by
direct numerical simulation of the soliton dynamics within the
framework of Eq. (\ref{E-eq}). Thus, we can conclude that, taking
into account the additional nonlinearity associated with the
second harmonics of the ion-ion hybrid frequency, in the radially
symmetric case there exists a stable two-dimensional ion-ion
hybrid soliton.

\section{\label{sec6} Conclusion}

We have derived a nonlinear system of equations for the envelope
of electrostatic potential at the ion-ion hybrid  frequency to
describe the interaction between an ion-ion hybrid waves and LF
acoustic-type disturbances in a magnetized plasma with two species
of ions. The resulting nonlinear equations also take into account
the contribution of second harmonics of the ion-ion hybrid
frequency. We have obtained a nonlinear dispersion relation
predicting the modulational instability of ion-ion hybrid waves.
For a number of particular cases, the modulational instability
growth rates have been found. By neglecting the contribution of
second harmonics, the phenomenon of collapse of ion-ion hybrid
waves is predicted. It has been also shown that taking into
account the interaction with the second harmonics suppresses
collapse of ion-ion hybrid waves and results in the existence of a
stable two-dimensional soliton. The developed theory is applicable
to a wide range of theoretical and experimental problems in both
space and laboratory (primarily devices with magnetic plasma
confinement) plasma with two species of ions.

A number of open questions remains to be addressed:

1) We have restricted ourselves to the 2D case, when the condition
(\ref{perp-cond}) is met and the ion-ion hybrid wave propagates
perpendicular to the external magnetic field. In a more general
case, it is necessary to take into account an additional term of
the form $\sim(k_{z}^{2}/k^{2})(m_{\alpha}/m_{e})$ in the
dispersion relation of the ion-ion hybrid wave Eq.
(\ref{disp-ion-hybr}). Then the model becomes three-dimensional
and essentially anisotropic. The anisotropy of the models in the
cases of upper- and lower-hybrid resonances results in the absence
of a stationary point of the Hamiltonian (for a fixed number of
plasmons), that is, in this case there are no three-dimensional
soliton solutions (even unstable ones)
\cite{KuznetsovZakharov2000,Scorich2010}. Note that the results
obtained in Refs.~\cite{KuznetsovZakharov2000,Scorich2010}
essentially use the cubic type of nonlinearity.  A similar
situation apparently occurs in the case of ion-ion hybrid
resonance.

2) The model under consideration takes into account the
interaction of HF ion-ion hybrid waves only with electrostatic LF
disturbances which corresponds to a perturbation of the plasma
density and neglects the interaction with nonpotential LF
disturbances of the Alfv\'{e}n type, which would correspond to an
LF perturbation of the magnetic field (the interaction of LF
Alfv\'{e}n waves with HF lower-hybrid and upper-hybrid waves was
considered in Refs.~\cite{Shukla2005,Lashkin2007}). Such neglect
corresponds to the smallness of the magnetic pressure in
comparison with the plasma gas-kinetic pressure, and is valid
under the condition $v_{A\alpha}\ll v_{s\alpha}$, where
$v_{A\alpha}=B_{0}/\sqrt{4\pi n_{0\alpha}m_{\alpha}}$ is the
Alfv\'{e}n velocity of the ion species $\alpha$.

3) Accounting for second harmonics is not the only reason for
stopping the collapse and the existence of stable 2D solitons. In
the static approximation, the Boltzmann distribution of electrons
and ions leads to a saturating exponential nonlinearity. Stable
multidimensional Langmuir solitons with this type of nonlinearity
were obtained in Refs.~\cite{Laedke1984,Lashkin2020}. Then,
apparently, the stable 2D ion-ion hybrid solitons could exist
without accounting for the contribution of second harmonics.

\section*{ACKNOWLEDGMENTS}

The author thanks O. K. Cheremnykh for discussions.

\appendix*
\section{}

\renewcommand{\theequation}{A\arabic{equation}}

In this appendix we write down the explicit expressions for the
coefficients $G_{1}$, $G_{2}$ and $G_{3}$ in Eq. (\ref{G-coef}),
\begin{gather}
G_{1}=[\nu_{2}T_{2}+\nu_{2} Z_{2}T_{e}(\nu_{1}
Z_{1}\Omega_{2}/\Omega_{1}+\nu_{2}
Z_{2})]\frac{(2\Omega_{1}^{2}-\omega_{ii}^{2})}{\omega_{ii}\Omega_{1}}
\nonumber \\
+[\nu_{1}T_{1}+\nu_{1} Z_{1}T_{e}(\nu_{1} Z_{1}+\nu_{2}
Z_{2}\Omega_{1}/\Omega_{2})]\frac{(2\Omega_{2}^{2}-\omega_{ii}^{2})}{\omega_{ii}\Omega_{2}},
\end{gather}
\begin{gather}
G_{2}=[\nu_{2}T_{2}+\nu_{2} Z_{2}T_{e}(\nu_{1} Z_{1}+\nu_{2}
Z_{2})]\frac{\Omega_{1}}{\omega_{ii}}
\nonumber \\
+[\nu_{1}T_{1}+\nu_{1} Z_{1}T_{e}(\nu_{1} Z_{1}+\nu_{2}
Z_{2})]\frac{\Omega_{2}}{\omega_{ii}},
\end{gather}
\begin{gather}
G_{3}=[\nu_{2}T_{2}+\nu_{2} Z_{2}T_{e}(\nu_{1} Z_{1}+\nu_{2}
Z_{2})]\frac{(\omega_{ii}^{2}-2\Omega_{1}^{2})}{\omega_{ii}\Omega_{1}}
\nonumber \\
+[\nu_{1}T_{1}+\nu_{1} Z_{1}T_{e}(\nu_{1} Z_{1}+\nu_{2}
Z_{2})]\frac{(\omega_{ii}^{2}-2\Omega_{2}^{2})}{\omega_{ii}\Omega_{2}},
\end{gather}
and also the coefficient $Q$ in Eq. (\ref{E-eq}),
\begin{gather}
Q=\frac{(\nu_{1} Z_{1}+\nu_{2}
Z_{2})(\omega^{2}_{p1}+\omega^{2}_{p2})^{2}G}{2\pi\varepsilon
(2\omega_{ii})e^{2}n_{0}(\Omega_{2}^{2}-\Omega_{1}^{2})^{2}R^{2}F}
\nonumber \\
\times
\left[\frac{(\Omega_{1}^{2}+\omega_{ii}^{2})}{Z_{1}(\Omega_{1}^{2}-4\omega_{ii}^{2})}+
\frac{(\Omega_{2}^{2}+\omega_{ii}^{2})}{Z_{2}(\Omega_{2}^{2}-4\omega_{ii}^{2})}\right].
\end{gather}

\end{document}